\documentstyle[preprint,floats,pra,aps,psfig]{revtex} 
\tightenlines  

\begin{document} \draft \textheight=9in \textwidth=6.5in 


\title{\LARGE \bf A single atom-based  generation of Bell states of two cavities } 

\author{A. Messina } 
\address{INFM, MURST and Dipartimento di Scienze Fisiche ed Astronomiche, via Archirafi 36, 90123 Palermo, Italy \\ Tel/Fax: +39 91 6234243; E-mail: messina@fisica.unipa.it }

\maketitle

\begin{abstract} 
A new conditional scheme for generating Bell states of two spatially separated high-Q cavities is reported. Our method is based on the passage of one atom only through the two cavities. A distinctive feature of our treatment is that it incorporates from the very beginning the unavoidable presence of fluctuations in the atom-cavity interaction times. The possibility of successfully implementing our proposal against cavity losses and atomic spontaneous decay is carefully discussed.
\end{abstract} 
  
\medskip


\def\shorttitle{Bell states of two cavities}

\pagebreak

The concept of entanglement is probably the most striking feature of quantum mechanics.
By definition a pure quantum state of two or more subsystems is said to be entangled if it is not a product of states of each component. 
It is worthnoting that while entanglement may be created
only if there exists a direct or indirect interaction mechanism
between the parts into play, generally speaking, an entangled state may describe a physical situation wherein the two or more entangled single 
subsystems are decoupled. 
The behaviour of the system in such a condition is dominated by the appearence of quantum correlations which become rather puzzling and counterintuitive when referred to well separated parts of the system.
Entanglement is the underlying mechanism for the measurement of quantum observables \cite{Peres} and is responsible for the occurrence of decoherence effects in the dynamics of quantum systems \cite{Zurek}.

In view of these considerations it is easy to convince oneself that it plays a central role in shading light on the problem of interface between the classical world and the quantistic one and, more in general, on the foundations of quantum mechanics itself \cite{Zurek}.
If, for example, the degree of correlation between two separate subsystems is large enough, Bell's inequalities can be violated \cite{Bell}. 
From an applicative point of view, quantum entanglement is the fundamental concept of the new field of quantum information providing a powerful physical source for a new kind of communication protocols \cite{Ekert,BEZ}.
Such protocols are, for example, essential in the quantum teleportation procedure which, on the other hand, is an extremely useful tool for understanding many properties of quantum entanglement itself.

The puzzling implications of entanglement as well as its applicability, have spurred an intense theoretical and experimental research work \cite{Gerry1,Grobe,Manzoor,Englert,Zheng,Molmer,Bergou}.

The recent developments in cavity quantum electrodynamics techniques as well as  the possibility of exciting and manipulating single Rydberg atoms, have provided favourable conditions for checking some puzzling quantum predictions related to the presence of entanglement. To this end, in particular, a great deal of attention has been devoted to the possibility of generating entangled states of spatially separated subsystems.
For example, many proposals for the generation of atomic entangled states have been presented \cite{Zheng,Molmer,Cirac,Bogar}.

As far as the radiation field state, a method for generating a Bell state  of two cavities has been reported, for example, as an intermediate step in the teleportation procedure proposed by Davidovich et al \cite{Davidovich}.
More recently, Zubairy et al \cite{Zubairy} have shown the possibility of teleporting a radiation field state from a cavity to another one, provided that the generation of a two-resonator entangled state for fixed number of photons inside the two cavities, is feasible.
Under the stimulus of these requirements in the context of teleportation procedures, theoretical schemes aimed at generating  entangled states of photons in many high-$Q$ resonators, have been very recently presented \cite{Manzoor,Bergou,Napoli,Meystre}.

In this paper we propose a simple and novel conditional method for the generation of Bell states of two spatially separated single-mode cavities.
The scheme we are going to discuss exploits the passage of a single atom only through the two cavities and the successive measurement of its internal state. We wish to underline from the beginning the relevance of this aspect from an experimental point of view. 
Preparing and controlling a single atom  is certainly much easier to achieve with respect to the case when the manipulation of many atoms is required.
In addition, taking into consideration the low efficiency $(\sim 50\%)$ \cite{HarocheQ} of the atomic state detectors today used in laboratory, conditional measurement procedures involving one atom only instead of many ones, have to be preferred. 
It is in fact easy to understand that the nonideal performance of these detectors drastically reduces the probability of success of multiatom conditional measurement schemes, eventually spoiling them of experimental interest.

In the context of CQED a common aspect of the proposals aimed at generating entangled states of two or more resonators is the requirement of ideal devices by which the interaction time between each crossing atom and the field may be sharply selected. On the other hand, the presence of unavoidable fluctuations in such interaction times might  invalidate the schemes themself, significantly reducing their practical interest. 
Bearing in mind this specific limitation of the experimental apparatus currently used in laboratory, we develop our theory incorporating in it, from the very beginning, the presence of fluctuations in the atom-cavity interaction times.

Another key condition for the effective generation of the Bell state of two spatially separated single-mode cavities, is the compatibility between the duration of the experiment and our ability to protect these states well enough against relaxation.
Concerning this delicate point, we check \textit{a posteriori} the possibility of successfully performing our experiment comparing, on the basis of published orders of magnitude for experimental parameters of interest, the time required for our preparation of Bell state with the cavity damping time. 

Let's start indicating by $\omega_1$ and $\omega_2$ $(\omega_1\sim \omega_2\sim 10^{10}Hz)$ the fundamental frequencies of the two cavities we wish to entangle. We shall demonstrate that, exploiting the passage of one atom only through the two resonators, it is possible to generate a Bell state of the two cavities.
To reach this goal, we use an effective three-level Rydberg atom whose three relevant states, and their relative energy levels, are respectively denoted by $\vert 0 \rangle$ $(E_0^A)$, $\vert 1 \rangle$ $(E_1^A)$ and $\vert 2 \rangle$ $(E_2^A)$  with $E_0^A<E_1^A<E_2^A$.
We impose the two resonance conditions $E_1^A-E_0^A\sim\omega_2$ and $E_2^A-E_1^A\sim\omega_1$   $(\hbar=1)$ and suppose that all the conditions under which the interaction between the atom and each cavity field can be cast in the form of the Jaynes-Cummings coupling, are satisfied.
The effective Hamiltonian describing the system under scrutiny, in the rotating wave approximation (RWA), can thus  be written down as 

\begin{equation}
H=H_0+H_1+H_2
\end{equation}
where

\begin{equation}
H_0=\sum^{2}_{i=1}{\omega_ia^\dagger_i a_i}+\sum_{j=0}^{2}{E_j^A\vert j\rangle \langle j \vert}
\end{equation}

\begin{equation}
H_1=g_1 a_1 \vert 2 \rangle \langle 1 \vert  +h.c. ;  \;\;\;\;\;\;
H_2=g_2 a_2 \vert 1 \rangle \langle 0 \vert   +h.c.
\end{equation}
In eqs. (2) and (3) $a_i(a_i^{\dagger})$ $(i=1,2)$ is the annihilation (creation) operator relative to the $i-$th cavity whereas  $g_1$ and $g_2$ measure the strenghts of the energy exchanges between the Rydberg atom and the cavity 1 and 2 respectively.
We point out that $g_i \not= 0$ $(i=1,2)$ only when the atom is inside the $i-$th cavity.

Suppose now that the atom prepared in the upper state $\vert 2 \rangle$ is injected into the cavity 1 previously excited in its $p-$photon state $\vert p \rangle_1$. After the interaction with the cavity 1 the atom enters the second cavity  also prepared in the state $\vert p \rangle_2$.
In order to include in our proposal the impossibility of sharply fixing the interaction times between the atom and the cavity 1 and 2 respectively, let's introduce the probability density $g_j(t,\tilde{t}_j)$ that $t$ is the
true duration of the interaction between the atom and the cavity $j (j=1,2)$.
We assume that $g_j(t,\tilde{t}_j)$ is a Gaussian distribution centered around $\tilde{t}_j$, hereafter referred to as the average interaction time between the atom and the $j$-th cavity, and having a spread $\Delta_j$ proportional to $\tilde{t}_j$. In other words we set 
\begin{equation}
g_j(t,\tilde{t}_j)=\frac{1}{\Delta_j \sqrt{2 \pi}}exp\{-\frac{(t-\tilde{t}_j)^2}{2 \Delta_j^2}\}
\end{equation}
where $\Delta_j=\gamma \tilde{t}_j$. The spread parameter $\gamma$ is related to the precision characterizing the method adopted for controlling the atom-field interaction time and its numerical value shall be specified later. Let's observe that letting $\Delta_j$ tends to zero, the probability density function $g_j(t,\tilde{t}_j)$ reduces to $\delta(t-\tilde{t}_j)$ thus describing the ideal control of the atom-field interaction time.

The occurrence of fluctuations in the duration of the atom-cavities coupling implies that, when the atom leaves the second resonator, the state of the coupled system (atom - cavity 1 - cavity 2) can be only described in terms of an appropriate density operator $\rho$.
Let's denote by $\vert \psi(t_1,t_2) \rangle$ the state of the atom-cavities system when the interaction times between the atom and the resonators 1 and 2 exactly coincides with $t_1$ and $t_2$ respectively. Observing that in a real experiment only the probability $(g_1(t_1,\tilde{t}_1)dt_1)(g_2(t_2,\tilde{t}_2)dt_2)$ of such an occurrence is indeed controllable, when the atom leaves the second cavity, the density operator $\rho$ describing the state of the combined atom-cavity fields may be represented as
\begin{equation}
\rho= \int_{-\infty}^{\infty}dt_2g_2(t_2,\tilde{t}_2)
\int_{-\infty}^{\infty}dt_1g_1(t_1,\tilde{t}_1)\vert \psi(t_1,t_2) \rangle \langle \psi(t_1,t_2)\vert
\end{equation}
where $\vert \psi(t_1,t_2) \rangle$ describes the state of the atom-cavities system when the interaction times between the atom and the resonators 1 and 2 exactly coincides with $t_1$ and $t_2$ respectively. 

Starting from the initial condition
\begin{equation}
\vert \psi(0) \rangle =\vert 2 \rangle \vert p \rangle_1
 \vert p \rangle_2
\end{equation}
and taking into account eqs. (2) and (3), it is not difficult to prove that 
\begin{eqnarray}
\vert \psi(t_1,t_2)\rangle&=&e^{-i\eta}\lbrace cos(g_1t_1\sqrt{p+1})\vert 2 \rangle\vert p \rangle_1 \vert p \rangle_2  \\ \nonumber
&-&icos(g_2t_2\sqrt{p+1})sin(g_1t_1\sqrt{p+1})\vert 1 \rangle\vert p +1\rangle_1 \vert p \rangle_2 \\ \nonumber
&-&sin(g_1t_1\sqrt{p+1})sin(g_2t_2\sqrt{p+1})\vert 0 \rangle\vert p+1 \rangle_1 \vert p+1 \rangle_2 \rbrace
\end{eqnarray}
where $\eta\equiv\eta(p,t_1,t_2)=[E_2^A+(\omega_1+\omega_2)p](t_1+t_2)$. 

Suppose now that after the interaction with both resonators the atom enters an appropriate Ramsey zone where the atomic ground and excited state, $\vert 0 \rangle$ and $\vert 2 \rangle$ respectively, are mixed. 
We wish to underline that, since the states $\vert 0 \rangle $ and $\vert 2 \rangle$ have the same parity, dipole transitions between them are forbidden. Thus, in order to produce a mixing between such two states 
either the selection rule of the levels parity is relaxed by a dc field or a multiphoton transition should be used \cite{referee}

Suppose that the interaction between the atom and the classical microwave field in the Ramsey zone is such that
\begin{equation}
\vert 0 \rangle \;\;\; \rightarrow \;\;\;
\frac{1}{\sqrt{2}}[\vert 0 \rangle +e^{i \chi}\vert 2 \rangle]; \;\;\;\;\;\;\;\;\;
\vert 2 \rangle \;\;\; \rightarrow \;\;\;
\frac{1}{\sqrt{2}}[\vert 2 \rangle -e^{-i \chi}\vert 0 \rangle] 
\end{equation}
where $\chi$ depends on the details of the atom-Ramsey zone field coupling.

When the atom leaves the Ramsey zone the density operator describing the state of the coupled system becomes

\begin{equation}
\rho_R= \int_{-\infty}^{\infty}dt_2g_2(t_2,\tilde{t}_2)
\int_{-\infty}^{\infty}dt_1g_1(t_1,\tilde{t}_1)\vert \psi_R(t_1,t_2) \rangle \langle \psi_R(t_1,t_2)\vert
\end{equation}
where
\begin{eqnarray}
\vert \psi_R(t_1,t_2)\rangle&=& 
e^{-i\eta}\lbrace\frac{1}{\sqrt{2}} cos(g_1t_1\sqrt{p+1})(\vert 2 \rangle-e^{i\chi}\vert 0\rangle )\vert p \rangle_1 \vert p \rangle_2 + \\ \nonumber
&-&icos(g_2t_2\sqrt{p+1})sin(g_1t_1\sqrt{p+1})\vert 1 \rangle\vert p+1\rangle_1 \vert p \rangle_2 \\ \nonumber
&-&\frac{1}{\sqrt{2}}sin(g_1t_1\sqrt{p+1})sin(g_2t_2\sqrt{p+1})(\vert 0\rangle+e^{i\chi}\vert 2\rangle )\vert p+1 \rangle_1 \vert p+1 \rangle_2 \rbrace
\end{eqnarray}
Inserting eq. (10) into eq. (9), we can conveniently rewrite  the density operator $\rho_R$ in the form:
\begin{eqnarray}
\rho_R&=& \frac{1}{2}\int_{-\infty}^{\infty}dt_2g_2(t_2,\tilde{t}_2)
\int_{-\infty}^{\infty}dt_1g_1(t_1,\tilde{t}_1)\vert \varphi^{(2)}_{field} \rangle \langle \varphi^{(2)}_{field} \vert   \vert 2 \rangle \langle 2 \vert  \\ \nonumber
&+& \frac{1}{2}\int_{-\infty}^{\infty}dt_2g_2(t_2,\tilde{t}_2)
\int_{-\infty}^{\infty}dt_1g_1(t_1,\tilde{t}_1) \vert \varphi^{(0)}_{field} \rangle \langle \varphi^{(0)}_{field} \vert   \vert 0 \rangle \langle 0 \vert  \\ \nonumber
&+& \rho_{residual}
\end{eqnarray}
where 
\begin{equation}
\vert \varphi^{(2)}_{field} \rangle=cos(g_1t_1 \sqrt{p+1})\vert p \rangle_1 \vert p \rangle_2-e^{i {\chi}}\prod_{i=1,2}{sin(g_it_i \sqrt{p+1})} \vert p+1 \rangle_1 \vert p+1 \rangle_2
\end{equation}
and
\begin{equation}
\vert \varphi^{(0)}_{field} \rangle=e^{-i {\chi}}cos(g_1t_1 \sqrt{p+1})\vert p \rangle_1 \vert p \rangle_2 +\prod_{i=1,2}{sin(g_it_i \sqrt{p+1})} \vert p+1 \rangle_1 \vert p+1 \rangle_2
\end{equation}
In eq. (11) the term $\rho_{residual}$ is characterized by the following properties:
\begin{equation}
\langle 0 \vert \rho_{residual} \vert 0 \rangle=\langle 2 \vert \rho_{residual} \vert 2 \rangle=0
\end{equation}
We shall not give its explicit form because, as we shall demonstrate later, in view of eq. (14), the terms contained in $\rho_{residual}$ are not involved in the generation procedure under scrutiny.
The last step of our scheme consists in measuring the atomic internal state by means an appropriate detector. As a result of this measurement the two cavities are projected onto an entangled state described by the following reduced density operator:
\begin{equation}
\rho_{field}=N_k \langle k \vert \rho_R \vert k \rangle
\end{equation}
provided that the atom is found in the state $\vert k \rangle$ $(k=0,2)$.
In eq. (15) the constant $N_k$ is evaluated  imposing $Tr \{ \rho_{field} \}=1$.

Our scheme is a conditional one in the sense that, fixing the Bell target state  to be produced, the experiment is considered realized with success if, and only if, the atom is measured in an appropriately prefixed internal state.
Let us observe that under ideal conditions, that is putting $\gamma=0$ in eq. (4), eq. (11) becomes
\begin{eqnarray}
\rho_R&=& \frac{1}{2}\vert \varphi^{(2)}_{field} \rangle \langle \varphi^{(2)}_{field} \vert   \vert 2 \rangle \langle 2 \vert  \\ \nonumber
&+& \frac{1}{2} \vert \varphi^{(0)}_{field} \rangle \langle \varphi^{(0)}_{field} \vert   \vert 0 \rangle \langle 0 \vert  \\ \nonumber
&+& \tilde{\rho}_{residual}
\end{eqnarray}
Looking at eqs.(16), (12) and (13) it is easy to convince oneself that in the ideal case, assuming that the atomic velocity can be manipulated in such a way that the conditions 
\begin{equation}
g_1t_1=\frac{\pi}{4 \sqrt{p+1}}; \;\;\;\; 
g_2t_2=\frac{\pi}{2 \sqrt{p+1}}\equiv 2g_1t_1
\end{equation}
are exactly satified, a measurement of the atomic internal state  produces the Bell state 
\begin{equation}
\vert \varphi^{(2)}_{field} \rangle_{id}=\frac{1}{\sqrt{2}}(\vert p \rangle_1 \vert p \rangle_2-e^{i \chi} \vert p+1 \rangle_1 \vert p+1 \rangle_2)
\end{equation}
if the atom is found in its upper state $\vert 2 \rangle$. On the other hand, the two cavities subsystem is projected onto its Bell state
\begin{equation}
\vert \varphi^{(0)}_{field} \rangle_{id}=\frac{1}{\sqrt{2}}(\vert p \rangle_1 \vert p \rangle_2+e^{i \chi} \vert p+1 \rangle_1 \vert p+1 \rangle_2)
\end{equation}
if the atom is measured in its ground state $\vert 0 \rangle$.
When, on the contrary, $\gamma \not=0$, as we have  said before, the state onto which the two cavities are projected as a result of the atomic internal state measurement, is given by 
\begin{equation}
\rho_{field}^{k}=N_k \int_{-\infty}^{\infty}dt_2g_2(t,\tilde{t}_2)
\int_{-\infty}^{\infty}dt_1g_1(t,\tilde{t}_1)\vert \varphi^{k}_{field} \rangle \langle \varphi^{k}_{field} \vert  
\end{equation}
provided that the atom is found in the state $\vert k \rangle$ $(k=0,2)$. 
Taking into account the results obtained under the hypothesis of ideal performance of the atomic velocity selector, 
we impose that the average interaction times between the atom and the cavities 1 and 2 , $\tilde{t}_1$ and $\tilde{t}_2$ respectively, satisfy the conditions (17).

In order to appreciate at what extent, with these choices for $\tilde{t}_1$ and $\tilde{t}_2$, eq. (20) describes a physical condition close to the ideal target states given by eqs. (18) and (19), we check the fidelity of our procedure evaluating the quantity
\begin{equation}
F(\gamma)=_{id} \langle \varphi^{(k)}_{field} \vert\rho_{field}^{(k)} \vert \varphi^{(k)}_{field} \rangle_{id}
\end{equation}
$F(\gamma)$ measures the realibility of our method of generating the Bell states $\vert \varphi^{(k)}_{field} \rangle_{id}$ given by eqs. (18) $(k=2)$ and (19) $(k=0)$ against  the presence of unavoidable fluctuations in the atom-cavity interaction times.
After lenghty calculations it is possible to demonstrate that 
\begin{equation}
F(\gamma)=\frac{2}{3+e^{-\frac{\gamma ^2 \pi ^2}{2}}} \{ \frac{1}{2}+ \frac{1}{4}
(1+e^{-\frac{\gamma ^2 \pi ^2}{2}})+e^{-\frac{\gamma ^2 \pi ^2}{4}} \}
\end{equation}
The function $F(\gamma)$ is plotted in figure (1) in the range of interest $0\div 0.1$. Its behaviour satisfactorily evidences that our procedure reaches its goal even in  the impossibility of sharply fixing the interaction time between the atom and each cavity.

The efficiency of the scheme we have presented is quantitatively expressed in terms of the probability $P_{(2)}(\gamma)$ or $P_{(0)}(\gamma)$ of finding the atom in its prefixed state $\vert 2 \rangle$ or $\vert 0 \rangle$ respectively after leaving the second cavity. It is easy to convince oneself that
\begin{equation}
P_{(k)}(\gamma)=\langle k \vert Tr_f (\rho) \vert k \rangle \;\;\;\;\;\;\;
k=0,2
\end{equation}
where $Tr_f$ means tracing over the field states.
Taking into consideration eq. (11), it is possible to prove that such a probability is independent from the result of the conditional measurement and can be written as:
\begin{equation}
P(\gamma) \equiv P_{(k)}(\gamma)=\frac{1}{4}(1+\frac{1}{2}(1+e^{-\frac{\gamma^2 \pi ^2}{2}}))\leq P(0)=\frac{1}{2} \;\;\;\; (k=0,2)
\end{equation}
Let's observe that $P(\gamma)$ is a decreasing function of $\gamma$ assuming its maximum value $\frac{1}{2}$ in correspondence to $\gamma=0$. 
The presence of interaction time fluctuations thus reduces the values of $P(\gamma)$ which, however, remains of experimental interest.

The fact that in our case the probability of success satisfies the condition $P(\gamma)\approx \frac{1}{2}$,  makes our single-atom scheme practically immune from the unavoidable negative consequences stemming from the low efficiency ($\approx 50 \%$) \cite{HarocheQ} of the atomic state measurement apparatus.

In our proposal the two interaction times between the atom and the two cavities are defined by eq. (17) so that, in general, they are different.
In order to practically realize this time sequence in laboratory, assume realistically $g_1$ and $g_2$ of the same order of magnitude and suppose to act upon the atomic velocity selector in such a way that the average time spent by the atom inside each cavity is just $\tilde{t}=\frac{\pi}{2g \sqrt{p+1}}\equiv t_2$, that is exactly coincident with the prefixed interaction time of the atom with the second cavity. Then the interaction time $t_1\equiv\frac{1}{2} \frac{g_2}{g_1}\tilde{t}$ between the atom and the first cavity, can be properly set tuning the atom in and out of resonance  through Stark field adjustment as described, for example, in ref. \cite{Davidovich}.

We wish now to discuss the experimental feasibility of our proposal taking into account other different possible important sources of nonideality.
To this end, it is of interest to note that, in the context of our proposal, it has been tacitly assumed that  both the cavity losses and the spontaneous atomic decay can be neglected.
In order to verify the extent of compatibility of these assumptions with the practical feasibility of our experimental scheme, we  compare the lifetime of the Rydberg atom, $\tau_A$, and the  damping time $\tau \simeq \frac{Q}{p\omega}$ $(p>1)$ of each cavity with the total duration $T$ of the experiment. 

To this end let's firstly note that in a typical experimental setup, the atom-field coupling constant $g$ ($g_1$ or $g_2$) can be chosen in such a way that $g\sim ( 10^5 \div 10^4) sec^{-1}$ \cite{WaltherQ}. Moreover the lifetime of a Rydberg atom is $\tau_A\geq 10^{-2}sec$ \cite{Haroche}.
As for the quality factor of the cavities currently used in laboratory, we note that the $Q$ value typical of the closed cavities used in the Munich experiments is $3 \times 10^{10}$ \cite{WaltherQ}.
On the other hand, experiments performed by the Paris group are based on open Fabry-Perot resonators characterized by quality factors which are about one order of magnitude smaller \cite{HarocheQ}.
Since our proposal requires the coherent mixing of two atomic states realized in an appropriate Ramsey zone just before the detection of the atomic internal state, open cavities have to be preferred with respect to the closed ones. The reason is that, due to the small exit holes ($3 mm$ diameter \cite{HarocheQ}) typical of the closed cavities, atomic coherences are spoiled by the presence of inhomogeneous stray of electric fields in these holes \cite{HarocheQ}.

The best cavities used in Paris experiments have a photon storage time $\tau\approx 1 ms$ corresponding to $Q\approx3 \times 10^{8}$ \cite{HarocheQ}.
The total duration $T$ of the experiment based on our proposal may be extimated as $T=\frac{l_{apparatus}}{v_A}$ where $l_{apparatus}$ is the linear size of the experimental set up and $v_A$ the atomic velocity. Considering that the lenght of each cavity is of the order of $1 cm$ \cite{HarocheR}, it is reasonable to assume $l_{apparatus}<10cm$. Moreover, we take $v_A=500\frac{m}{s}$ \cite{Haroche01}.
Taking into account these realistic experimental values, we immediately get $T\approx 2\times 10^{-4}sec$ which turns out to be satisfactorily larger than $\tau_A$.
Requiring $T< \tau$ yields 

\begin{equation}
T\approx 2\times 10^{-4}sec< \tau \approx \frac{Q}{p \omega}\approx 
\frac{3 \times 10^{8}}{p 10^{10}}sec=\frac{1}{p}10^{-2} sec 
\end{equation}
which shows that we may be confident with the fact that the effects due to cavity losses may be neglected, provided that the number of photons initially stored in each cavity is small.
With reference to this last point, it is worth noting that our theory might be tested in laboratory considering that Fock states with one or two photons have been realized \cite{Fockstate}.

We however emphasize that in simulations of the maser operation, Fock states up to $p=5$ can be readily generated for achievable experimental condition \cite{WaltherQ}.

We wish to conclude this paper pointing out that, in principle, it is possible to address the characterization of the produced entanglement between the two spatially separated cavities. To this end, we simply follow an idea suggested by Meystre \cite{Meystre}.
Suppose that a two-level probe atom interacting with the first cavity only, is prepared in its ground state and sent through the two resonators at the end of our experiment.
It is not difficult to persuade oneself that the probability of detecting the atom in its initial state at the exit from the second cavity, is an oscillating time function depending on $p$ and on the quantum coherence $\chi$.
This circumstance gives, in principle, the possibility to distinguish our Bell superpositions from the correspondent mixture and/or to experimentally prove the existence of the desired two-cavity entangled state.

\pagebreak
{\LARGE \bf Acknowledgments} 

 The author expresses his gratitude to Dr. A. Napoli for reading the manuscript and for stimulating conversations on the subject of this paper.


\pagebreak

Figure Caption

FIGURE 1: Fidelity of the scheme $F(\gamma)$ in the range of interest $0 \div 0.1$.

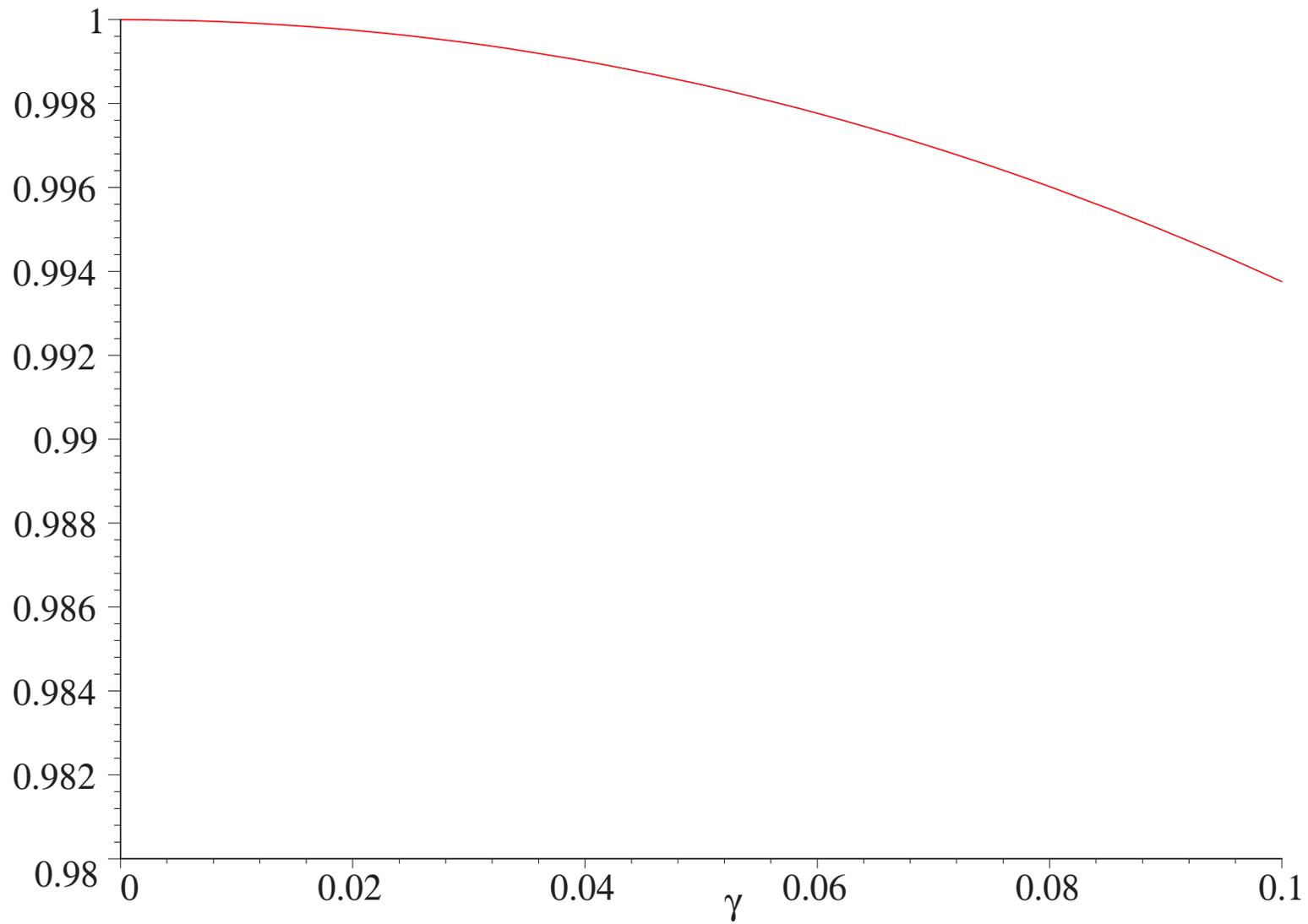
\end{document}